\documentclass[a4paper,11pt]{article}
\pdfoutput=1
\usepackage{jcappub}

\usepackage{amsmath}
\usepackage{bm}
\usepackage{xcolor,color,colortbl}
\usepackage[utf8]{inputenc}
\usepackage{hyperref}
\usepackage{url}
\usepackage{graphicx}
\usepackage{multirow,bigdelim}
\usepackage{amssymb}
\usepackage[amssymb]{SIunits}
\usepackage{multirow}
\usepackage{bbold}
\usepackage{verbatim}
\usepackage{comment}
\usepackage{caption}
\usepackage{physics}
\usepackage{subcaption}
\usepackage{orcidlink}

\usepackage{tikz}
\usetikzlibrary{positioning}

\definecolor{darkgreen}{HTML}{339933}

\usepackage{scalerel,tikz}
\usetikzlibrary{svg.path}
\definecolor{orcidlogocol}{HTML}{A6CE39}
\tikzset{orcidlogo/.pic={
 \fill[orcidlogocol] svg{M256,128c0,70.7-57.3,128-128,128C57.3,256,0,198.7,0,128C0,57.3,57.3,0,128,0C198.7,0,256,57.3,256,128z};
 \fill[white] svg{M86.3,186.2H70.9V79.1h15.4v48.4V186.2z}
 svg{M108.9,79.1h41.6c39.6,0,57,28.3,57,53.6c0,27.5-21.5,53.6-56.8,53.6h-41.8V79.1z M124.3,172.4h24.5c34.9,0,42.9-26.5,42.9-39.7c0-21.5-13.7-39.7-43.7-39.7h-23.7V172.4z}
 svg{M88.7,56.8c0,5.5-4.5,10.1-10.1,10.1c-5.6,0-10.1-4.6-10.1-10.1c0-5.6,4.5-10.1,10.1-10.1C84.2,46.7,88.7,51.3,88.7,56.8z};
}}
\newcommand\orcidicon[1]{\href{https://orcid.org/#1}{\mbox{\scalerel*{
\begin{tikzpicture}[yscale=-1,transform shape]
\pic{orcidlogo};
\end{tikzpicture}
}{|}}}}

\title{The One-Loop Power Spectrum of Fast Radio Burst Dispersion Measures}

\author[a,b,c]{Haruki Ebina~\orcidicon{0000-0002-1080-0955}}
\author[a,b,c]{and Martin White~\orcidicon{0000-0001-9912-5070}}
\affiliation[a]{Department of Physics, University of California, Berkeley, CA 94720, USA}
\affiliation[b]{Berkeley Center for Cosmological Physics, UC Berkeley, CA 94720, USA}
\affiliation[c]{Lawrence Berkeley National Laboratory, One Cyclotron Road, Berkeley, CA 94720, USA}

\emailAdd{ebina@berkeley.edu}

\abstract{
Fast radio burst (FRB) dispersion measures trace free electron column densities and will soon offer a unique probe of low-$z$ baryons. 
Such measurements can constrain cosmology directly, with the free electrons serving as a new tracer of large-scale structure, and probe the baryonic feedback of galaxies, a leading systematic for weak lensing surveys such as LSST and Euclid.
We prepare for both science cases using effective field theory (EFT) and hydrodynamical simulations. We construct the one-loop EFT description of the free-electron auto-spectrum $P_{ee}$ and electron-galaxy cross-spectrum $P_{eg}$, placing FRB dispersion clustering on the same theoretical footing as spectroscopic galaxy analyses, and quantify the FRB densities at which this modeling is useful. We also investigate suitable galaxy samples for cross-correlations, finding that current spectroscopic catalogs provide appropriate redshift range and sufficient density.
We validate the model against the FLAMINGO simulations, jointly fitting $P_{ee}$, $P_{eg}$, and $P_{gg}$ for DESI-like samples at $z=0.2$ and 0.5. The model describes all three spectra to $k\sim0.2\,h\,{\rm Mpc}^{-1}$, with an electron linear bias $b_{e,1}\simeq0.92$, higher-order biases consistent with zero, and all parameters stable across feedback variants. 
Together with the near-perfect electron-matter correlation $r_{em}\simeq1$, this establishes free electrons as nearly unbiased, feedback-robust tracers of matter, supporting a key assumption of FRB-based feedback constraints. These properties make electron clustering an ideal application for Hybrid Effective Field Theory (HEFT), which would extend the modeling reach by a further factor of 2--3. The low-$z$ electron spectrum becomes signal-dominated beyond the linear regime within the first few years of next-generation surveys; the models developed here will be necessary on these timescales.
}

\begin{document}
\maketitle
\flushbottom

\section{Introduction}

Fast radio burst (FRB \cite{Bhandari21,Petroff22}) dispersion measures, the frequency-dependent delays in the radio signal, trace the column density of free electrons along the line of sight to the burst.  At $z\lesssim3$, free electrons reside overwhelmingly in the diffuse ionized phase that hosts nearly 90\% of the baryons at low redshift \cite{McQuinn14,Peroux20} and trace the underlying baryon field at the few-percent level \cite{Leung25}. This opens the way to a new, independent probe of baryons and novel constraints on the complex astrophysics associated with galaxy feedback \cite{Ebina26e}. FRBs thus provide critical new information on both galaxy formation and baryonic feedback, one of the leading systematics for Stage-IV weak lensing cosmology surveys such as LSST \cite{LSST} and Euclid \cite{Euclid}. The latter relies on the assumption that the free electrons trace the matter field very closely, which we will confirm in this work using hydrodynamical simulations. 
Simultaneously, the free electrons measured through FRBs can be modeled as a new (potentially) biased tracer of the underlying matter, allowing for an independent probe of large-scale structure cosmology \cite{Dodelson20}. 

These science cases are supported by the rapidly growing ability to detect and localize FRBs. Current-generation surveys, such as CHIME (Canadian Hydrogen Intensity Mapping Experiment) \cite{CHIME} (through their outrigger program \cite{Lanman24}), ASKAP (Australian SKA Pathfinder) \cite{askap}, and DSA-110 (Deep Synoptic Array 110) \cite{Law24}, are capable of detecting $\mathcal{O}(100)$ localized FRBs per year \cite{Lanman24,JahnsSchindler23}. Next-generation surveys, like DSA \cite{Hallinan19,Connor26}, CHORD (Canadian Hydrogen Observatory and Radio-transient Detector) \cite{Vanderlinde19}, and SKA (Square Kilometre Array) \cite{Macquart15,Zhang23} are expected to detect $\mathcal{O}(10^4)$ localized FRBs per year \cite{Hallinan19,Vanderlinde19}; both DSA and CHORD are planned for first light in 2028. 

The integrated nature of the dispersion signal, and the often uncertain source redshifts, suggest that cross-correlation with a galaxy sample of well-known redshift distribution would be a powerful method for signal localization in redshift and systematics mitigation. Galaxy catalogs with spectroscopic redshifts are ideal candidates for this cross-correlation, with the spectroscopic redshifts ensuring full characterization of the (often narrow) catalog redshift distributions. 
Spectroscopic redshift surveys, such as the Dark Energy Spectroscopic Instrument (DESI) \cite{DESI,DESI-DR1}, are one of the leading methods to conduct large-scale structure cosmology and they achieve this by spectroscopically obtaining redshifts of tens of millions of galaxies, mapping the density field in the dark- energy- and matter-dominated era in 3D. 
This allows them to measure the power spectra at sub-percent precision and has recently revealed some tension with the currently standing $\Lambda$CDM model, inferring potential for new physics \cite{DESI24-V,DESI-DR2}. The data collection of spectroscopic surveys through DESI and other concurrent surveys \cite{Euclid,PFS} will continue rapidly, allowing for increased measurement precision over time. 
The ability to extract information from cross-correlation with large, well-characterized, and expanding galaxy catalogs brings an opportunity for FRB cosmology. However, in order to model this cross-correlation correctly and to bring FRB measurements to the same standing as the large-scale structure surveys working at unprecedented precision, it is prudent to have a consistent modeling of FRB dispersions as that of galaxies and other large-scale structure tracers. 

The current generation of cosmological hydrodynamic simulations predicts that on large scales these electrons can be modeled with linear perturbation theory and as biased tracers of the matter density field with a bias very close to unity \cite{Masui15,Zhou25b,Andrew26}. 
Forecasts indicate that next-generation FRB experiments will yield competitive cosmological constraints, primarily by employing FRBs alone or in cross-correlation with galaxy catalogs under linear theory (e.g.~\cite{Masui15,Madhavacheril19,Sharma26,Raikman26}). 
The one-loop perturbation model developed in this work extends these analyses to smaller scales, which drastically increases the number of available Fourier modes in analyses, and enables a seamless integration of FRBs with galaxy surveys with the same model extent and rigor as their key analyses \cite{DESI24-V}. This can further be extended by using Hybrid Effective Field Theory (HEFT), which combines N-body simulations with perturbative models to increase the small-scale limit by another factor of $\sim2$-3. 
We will confirm the perturbation theory models using all feedback models available in the FLAMINGO suite of hydrodynamical simulations, finding $b_{e,1}\simeq0.92$ and higher-order biases consistent with zero across model variants. In addition, we quantify the extent to which free electrons trace the underlying matter, which is an essential component of using FRBs to model baryonic feedback. 

This work is organized as follows. First, in \S\ref{sec:frbintro}, we introduce the FRB dispersion measures as a large-scale structure probe, and seek insight into the requirements on the FRB catalogs, galaxy samples, and theoretical modeling for next-generation experiments. Next in \S\ref{sec:formalism}, we use the one-loop effective field theory formalism to model the FRB dispersion auto-correlation and dispersion-galaxy cross-correlation on the same footing as galaxy surveys. In \S\ref{sec:heft} we identify the possibility to extend the theoretical description for FRB clustering, using Hybrid Effective Field Theory (HEFT). 
We then test and confirm this formalism in \S\ref{sec:sim} using hydrodynamical simulations with various feedback models. These simulations will also confirm for us how closely free electrons trace matter, an essential requirement for science cases like baryonic feedback. Finally, we conclude in \S\ref{sec:conclusion}.

\section{Fast Radio Burst Dispersion Measure}
\label{sec:frbintro}

\subsection{The Dispersion Measure}

FRBs, through the frequency dependence of the dispersion measure (DM), trace the integrated density of free electrons in the diffuse, ionized gas, which hosts ${\sim}90\%$ of cosmic baryons at low redshift \cite{Peroux20}. This can be expressed as\footnote{The $(1+z)$ scaling in Eq.~\ref{eq:D} arises from a competition of multiple effects.  As $D\propto \Delta t_{\rm obs}\,\nu_{\rm obs}^2$, time dilation and frequency redshifting require $D\propto n_e\,d\ell/(1+z)$ with proper lengths and densities.  Converting to comoving distances and densities gives the $1+z$ in Eq.~\ref{eq:D}.}
\begin{equation}
    D(\hat{\mathbf{n}}) =  \int_0^{\chi_f} d\chi \  \bar{n}_e^{\rm com}(z)\,(1+z)(1 + \delta_e(\hat{\mathbf{n}}, z))
    \label{eq:D}
\end{equation}
where $\hat{\mathbf{n}}$ is the direction of the line of sight, $\chi$ is the comoving distance, $\chi_f$ is the comoving distance to the FRB, $z$ is the redshift corresponding to $\chi$, $\delta_e(\hat{\mathbf{n}}, z)$ is the overdensity of free electrons at redshift $z$, and $\bar{n}_e^{\rm com}(z)$ is the comoving, mean free electron density
\begin{equation}
    \bar{n}_e^{\rm com}(z) = \Omega_b\, \rho_{{\rm cr},0}\, f_e(z) \, m_p^{-1} (1 - Y/2) \quad ,
\end{equation}
where $\rho_{{\rm cr},0} \equiv 3H_0^2/8\pi G$ is the present-day critical density, $\Omega_b \equiv \rho_{b,0}/\rho_{\mathrm{cr},0}$ is the present-day baryon density, $Y$ is the primordial helium mass fraction, $m_p$ is the proton mass, and $f_e(z) \equiv \bar{\rho}_e(z)/\bar{\rho}_b(z)$ is the fraction of baryons residing in diffuse ionized gas at redshift $z$ \cite{McQuinn14,Masui15,Shirasaki17,Macquart20,Zhou25b,Andrew26}. 
This decomposition makes explicit the dependence of the DM signal on $\Omega_b$, while the astrophysical uncertainties, such as the effects of feedback processes that redistribute baryons between diffuse and collapsed phases, are absorbed into $f_e(z)$.
If $f_e(z)=1$ then $\bar{n}_e$ is independent of redshift.

As DM ($D$) is an integrated measure, one cannot measure the 3D clustering directly. Instead, one can analyze the two-point correlation of these fields, e.g.\ through their angular power spectrum\footnote{For FRBs with a distribution of redshifts this expression needs to be further integrated over $dN_{\rm FRB}/dz$.}
\begin{equation}
    C_\ell^{DD} = \bar{n}_{e,0}^2 \int_0^{\chi_f} d\chi \, \frac{(1+z)^2}{\chi^2} \, \left(\frac{f_e(z)}{f_{e,0}}\right)^2 P_{ee}\left(k=\frac{\ell+1/2}{\chi}, z\right)
\end{equation}
where we apply the extended Limber approximation for the projection of $P_{ee}$ \cite{Limber,LovAfs08,Wayland26}, $\bar{n}_{e,0}=\bar{n}_e(z=0)$, $f_{e,0}=f_e(z=0)$, and at $z<1$, $f_e(z)/f_{e,0}\approx 1$ \cite{Leung25,Zhou25b,Andrew26}. 
Direct detection and analysis of this signal are challenging due to three aspects of FRBs. First, the line-of-sight integration to the source dilutes the SNR as $\sim1/\chi_f$. Second, a competitive measurement requires an abundant number of FRBs that are localized at a sufficiently narrow source redshift. Third, it is potentially subject to host-galaxy dispersion correlations that are not well characterized. 
The first two issues, if not all three, can be overcome in time as the FRB catalog grows. Cross-correlation with galaxy populations (with sufficiently narrow redshift distributions), however, yields tighter constraints on a far shorter timescale, and may be intrinsically advantageous being largely immune to host-galaxy correlations.
Given a galaxy population with a narrow redshift distribution around redshift $z_g$, the cross-spectrum is
\begin{equation}
    C_\ell^{Dg} = \bar{n}_{e}(z_g) \, \frac{1 + z_g}{\chi_g^2} \, P_{ge}\left(k=\frac{\ell+1/2}{\chi_g}, z_g\right)
\end{equation}
where $\chi_g$ is the corresponding comoving distance and $P_{ge}(k, z_g)$ is the three-dimensional cross-power spectrum between the galaxy overdensity field, $\delta_g$, and the free electron overdensity field, $\delta_e$, evaluated at redshift $z_g$ \cite{Shirasaki17,Madhavacheril19}. We refer the reader to refs.~\cite{Rafiei-Ravandi20,Alonso21,Sharma26,Wayland26} for expressions and derivations of $C_\ell^{Dg}$ involving a general (not thin-shell) population of galaxies and FRBs. 
This signal notably circumvents the three major caveats of the auto-spectrum $C_\ell^{DD}$ discussed above, by making use of all FRBs with redshift sufficiently larger than $z_g$ (in order to avoid FRB source-galaxy correlation), by localizing the $\chi$-integral to the galaxy population, and by excluding host-host correlations. In addition, the precision requirement for the FRB host-halo redshift is significantly relaxed. Localizing the $\chi$-integral also tightens the mapping between $k$ and $\ell$, which simplifies the modeling and allows distinct physical scales to be separated in the observed spectrum.

\begin{figure}
    \centering
    \includegraphics[width=\linewidth]{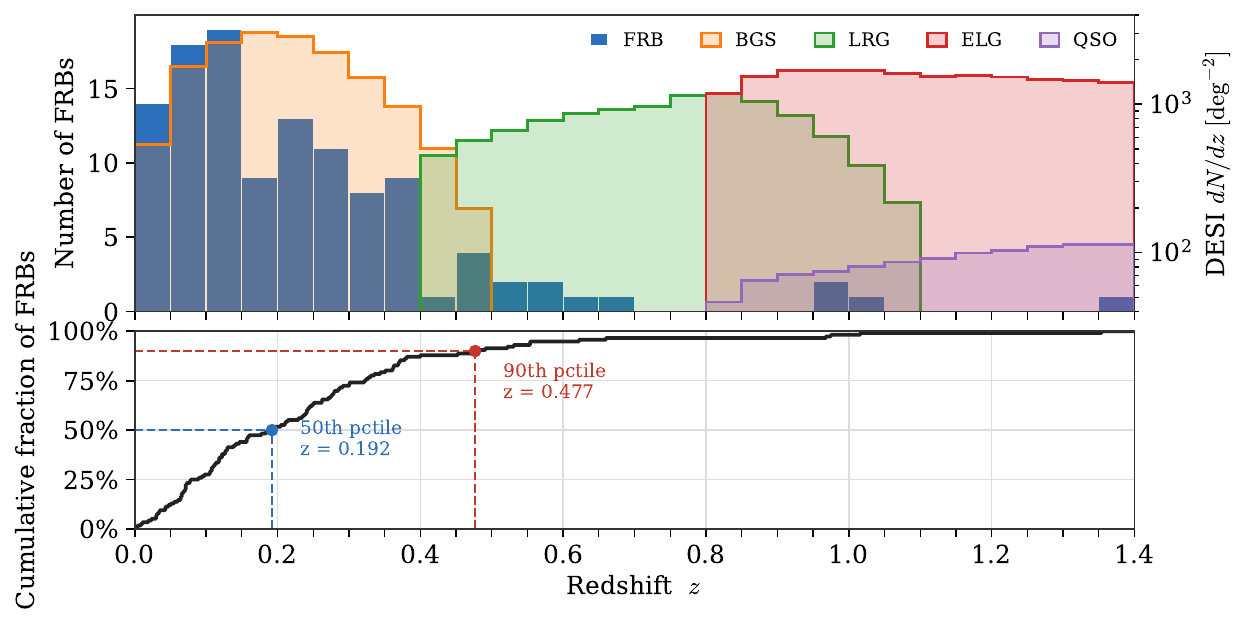}
    \caption{Redshift distribution of currently localized FRBs, shown alongside the DESI tracer samples: the Bright Galaxy Survey (BGS), Luminous Red Galaxies (LRG), Emission Line Galaxies (ELG), and quasars (QSO). The FRB distribution indicates that $\sim50$\% of localized FRBs are above $z=0.2$, whereas only $\sim10$\% are above $z=0.5$, illustrating the difficulty of performing cross-correlations at higher redshifts. At $z=0.2$ and 0.5, the spectroscopic galaxies of primary interest for cross-correlation will be BGS and LRG, respectively.}
    \label{fig:FRBz}
\end{figure}

\subsection{Current Data and Future Landscape}

The current database and future forecast of FRBs illuminate the feasibility of FRB-galaxy cross-correlations, and the necessary requirements on the galaxy catalog and theoretical modeling of the power spectra ($P_{ee}$, $P_{eg}$, $P_{gg}$). 
At the moment, there are $\gtrsim100$ localized FRBs, 116 of which we collect \cite{Chatterjee17,Mahony18,Bannister19,Prochaska19,Ravi19,Bhandari20,Heintz20,Law20,Marcote20,Bhardwaj21a,Bhardwaj21b,Chittidi21,Day21,Bhandari22,Ocker22,Rajwade22,Bhandari23,Caleb23,Chime23,Michilli23,Ravi23,Ryder23,Shannon23,Baptista24,Bhardwaj24,Cassanelli24,Faber24,Law24,Rajwade24,Tian24,Connor25,Chime25,Gao25,Shah25,Shannon25} from the blinkverse database\footnote{117 FRBs are available at the blinkverse database, from which we exclude FRB 20190614 due to having no spectroscopic redshift and being associated with more than one host halo. Furthermore we assign $z=0$ to FRB 20200120E, which is a local structure. 106 of 117 localized FRBs can be found in Table 6 of ref.~\cite{Lemos26}.} \cite{Xu23}. In the near-term we are FRB-limited, rather than galaxy-limited, driving cross-correlation efforts to lower redshift galaxies with as many FRBs behind the galaxies (at higher redshift) as possible. This is illustrated in Fig.~\ref{fig:FRBz}, showing the overlap in redshift range between source FRBs and DESI galaxies and the rapid decay of FRBs as one moves to higher redshift. 
Note that this simple diagnostic implicitly assumes that the FRB redshift distribution stays consistent into next-generation experiments, which is likely not the case. Taking the detection capability of next-generation experiments into account, the redshift distribution of detected FRBs are likely to shift to higher redshift, peaking around $z\sim0.6$ with the high-$z$ distribution tail extending past $z\sim2$ \cite{Feng26}. Even in this case, the majority of FRBs remain covered by DESI galaxies ($z_g\lesssim1.6$)\footnote{If one were interested in cross-correlating with higher-redshift galaxy tracers, one could employ sparser quasar samples extending to $z\sim3$ \cite{DESI-DR2}, or the galaxy tracers planned for the next-generation spectroscopic redshift surveys \cite{Wilson19,Ferraro22,Schlegel22,Beseuner25}. These tracers have already been targeted, observed spectroscopically, and analyzed in early clustering studies aimed at establishing their cosmological utility \cite{Ruhlmann-Kleider24,Payerne24,Ebina26a,Ebina26c}.}, especially accounting for the requirement $z_{\rm FRB}>z_g$ for cross-correlation, and it remains true that low-$z$ cross-correlation benefits from more background FRBs. 

\begin{figure}
    \centering
    \includegraphics[width=\linewidth]{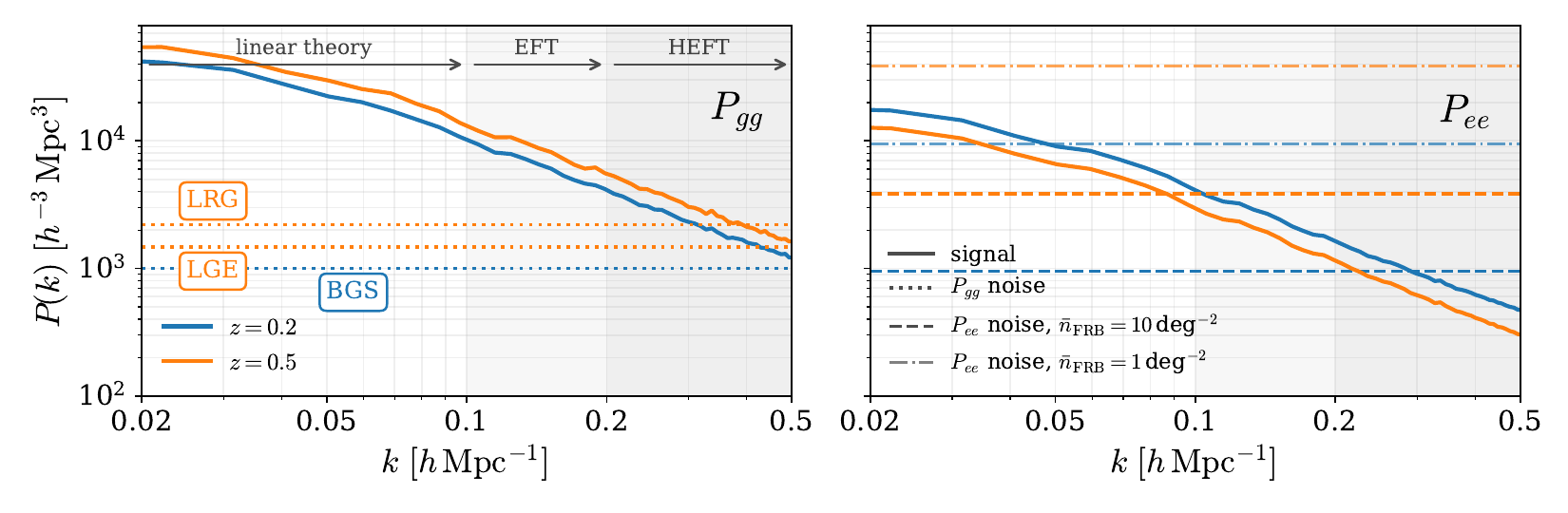}
    \caption{The signal ($P$, excluding shot noise as measured from simulations in \S\ref{sec:sim}) against the noise for galaxy (left) and free electrons (right) probed by FRB-galaxy cross-correlations, assuming $\sigma_D=100\,\mathrm{pc}\,\mathrm{cm}^{-3}$, DESI-like galaxies (see \S\ref{sec:sim} and refs.~\cite{DESI24-II,DESI-DR2}) and $\Delta z=0.3$. The $z=0.2$ and 0.5 samples are shown in blue and orange colors, respectively. The signal are shown in solid lines, while noise are shown in dotted lines for the galaxies, and dash-dot and dashed line for the FRBs at $\bar{n}_{\rm FRB}=1\,\deg^{-2}$ and $\bar{n}_{\rm FRB}=10\,\deg^{-2}$, respectively. The two orange dotted lines on the left panel show the current (LRG) and soon expected (LGE) galaxy population densities at $z=0.5$.
    The near-term FRB density $\bar{n}_{\rm FRB}=1\,\deg^{-2}$ indicates noise domination for electrons for most of the dataset, save the largest scales at $z=0.2$. The $\bar{n}_{\rm FRB}=10\,\deg^{-2}$ FRBs in the longer term significantly changes the picture, with the $z=0.2$ sample pushing past the linear theory regime ($k\sim0.1\,h\,\mathrm{Mpc}^{-1}$). In both cases, the signal-noise equilibrium for electrons arrive significantly before that of the galaxies, indicating that constraint improvements from increased galaxy density are limited. 
    Note that $\bar{n}_{\rm FRB}$ shows the density of background FRBs, hence achieving the same $\bar{n}_{\rm FRB}$ is harder at higher redshift, as shown in Fig.~\ref{fig:FRBz}.}
    \label{fig:snr}
\end{figure}

\begin{figure}
    \centering
    \includegraphics[width=\linewidth]{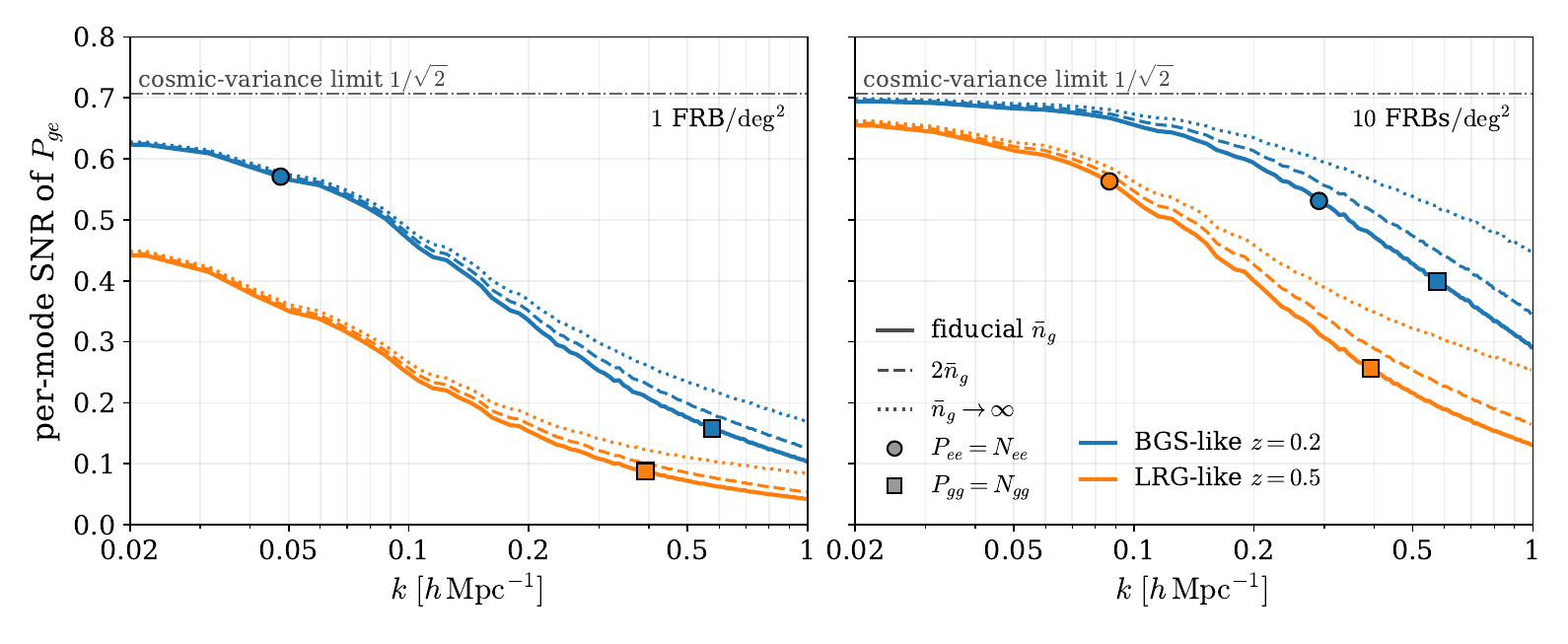}
    \caption{The per-mode SNR of $P_{ge}$ given the FRB densities $\bar{n}_{\rm FRB}=1$ (left) and $10\,\deg^{-2}$ (right), assuming $\sigma_D=100\,\mathrm{pc}\,\mathrm{cm}^{-3}$, $\Delta z=0.3$, and DESI BGS (blue, at $z=0.2$) and LRG (orange, at $z=0.5$) like galaxies. 
    The different line styles demonstrate fiducial (solid), double (dashed), and infinite (dotted) galaxy densities. 
    The solid points refer to the scale at which galaxy (square) and electron (circle) spectra reach SNR of unity per mode. 
    Due to the cross-term of the $P_{ge}$ variance, even in electron noise dominated regimes an increase in galaxy density leads to a gradual increase in SNR. 
    At $k=0.6\,h\,\mathrm{Mpc}^{-1}$, the $z=0.2$ and 0.5 samples increase SNR by 15\% (13\%) and 21\% (20\%), respectively, with $\bar{n}_{\rm FRB}=1$ (10) $\deg^{-2}$.}
    \label{fig:SNR_Pge}
\end{figure}

In the longer term we expect substantially more FRBs, with $\mathcal{O}(10^4)$ FRBs per year from next-generation experiments starting in the next few years \cite{Hallinan19,Vanderlinde19}. With a sky coverage of $f_\mathrm{sky}\sim0.5$, this equates to $\mathcal{O}(1\,\deg^{-2})$ FRBs per year and $\mathcal{O}(10\,\deg^{-2})$ by the end of the surveys. This allows us to estimate what galaxy populations and theoretical readiness are required to model FRB-galaxy cross-correlations to its full potential. The observed DM field contains both the cosmological signal $D_{\rm sig}$ and a noise term $\epsilon$,
\begin{equation}
D(\hat{\mathbf{n}}) = D_{\rm sig}(\hat{\mathbf{n}}) + \epsilon(\hat{\mathbf{n}}),
\end{equation}
where $\epsilon$ is assumed uncorrelated from burst to burst, with contributions from measurement error, the stochastic host-galaxy and circumburst plasma, and residual errors in the Milky Way subtraction. Assuming $\epsilon$ is uncorrelated between bursts its power spectrum is white,
\begin{equation}
N_\ell^{DD} = \frac{\sigma_{D}^2}{\bar{n}_{\rm FRB}},
\label{eqn:NellDD}
\end{equation}
where $\bar{n}_{\rm FRB}$ is the angular number density of FRBs and $\sigma_D$ is the per-burst rms of $\epsilon$. For cosmological FRBs, $\epsilon$ is expected to be dominated by the host and circumburst contribution. Recent fits to localized samples motivates $\sigma_{D} \approx 100\,\mathrm{pc\,cm^{-3}}$ \cite{Connor25}, and this value has been adopted in recent forecasts \cite{Sharma26,Andrew26}, although the inferred scatter carries sizeable uncertainties and depends on the FRB sample. We shall fix $\sigma_{D}=100\,\mathrm{pc\,cm^{-3}}$ and interpret $\bar{n}_{\mathrm{FRB}}$ in Eq.~\ref{eqn:NellDD} as an `effective' number density (as is done in cosmic shear studies).
At a thin redshift slice around comoving distance $\chi$, this translates into 3D noise of the electron power spectrum of
\begin{equation}
    N_\ell^{DD} = N_{ee} \,\Delta\chi \frac{W_D^2(\chi)}{\chi^2}
    \label{eq:Nee}
\end{equation}
where $W_D=\bar{n}_{e}(z)(1+z)$ implicitly encodes cosmology dependence through $\bar{n}_{e}$. 
By comparing this against the galaxy shot noise $N_\ell^{gg}=1/\bar{n}^{2D}_g$, we are able to derive the sufficient galaxy population for cross-correlation; this is shown in Fig.~\ref{fig:snr}, for $z=0.2$ and 0.5 with slices of $\Delta z=0.3$, using simulation spectra from \S\ref{sec:sim}. 
A comparison between the galaxy and electron panels indicates that in all scenarios, the cross-correlation uncertainty is dominated by the electron noise rather than the galaxy noise, indicating that current DESI-like spectroscopic samples therefore already provide both the redshift coverage (Fig.~\ref{fig:FRBz}) and the number density required for these cross-correlations. 
Due to the structure of the cross-spectrum covariance ($\mathrm{Var[P_{eg}]}\propto [P_{gg}+N_{gg}][P_{ee}+N_{ee}] + [P_{eg}+N_{eg}]^2$), an increased galaxy number density, such as that  expected from the Luminous Galaxy Extension (LGE) sample in DESI Run 1b (2026-2028), may still increase the SNR by suppressing the effect of DM noise at `high' $\ell$. In the noise-dominated regime of electrons ($\mathrm{Var[P_{eg}]}\approx [P_{gg}+N_{gg}]N_{ee}$), at scales where $P_{gg}\sim N_{gg}$, doubling the galaxy density would improve the SNR by $\sqrt{4/3}\approx1.15$, as shown in Fig.~\ref{fig:SNR_Pge}\footnote{The Extreme Emission Line Galaxies (XLG) sample of DESI Run 2, planned to start in 2029, will not only substantially raise number densities at $0.2\lesssim z\lesssim0.9$, but also yield low-density tracers \cite{Schlegel22,Beseuner25}. The significant change in bias has the potential to qualitatively change the situation and must be assessed independently.}.
While these gains are non-negligible, pursuing them at the expense of spectroscopic redshifts is likely not a positive trade. Photometric redshifts would restrict the galaxy auto-correlation to the projected angular power spectrum $C_\ell^{gg}$, forfeiting the three-dimensional $P_{gg}$ and with it much of the constraining power on the galaxy biases.

The electron noise dominates on most scales in the near term ($\bar{n}_{\rm FRB}=1\,\deg^{-2}$), while in the longer term ($\bar{n}_{\rm FRB}=10\,\deg^{-2}$), signal dominates to $k\sim0.3\,h\,\mathrm{Mpc}^{-1}$ and $0.1\,h\,\mathrm{Mpc}^{-1}$ for $z=0.2$ and 0.5, respectively, although both of these are technically subject to the uncertainty of $\sigma_D$. 
Specifically, for FRB signal domination to $k=0.1$ (`lin') and 0.2$\,h\,\mathrm{Mpc}^{-1}$ (`EFT') respectively, one would require
\begin{align}
    N^{DD}_{\ell,\rm lin} = 4300\, \left(\frac{\mathrm{pc}}{\mathrm{cm}^3}\right)^2 \deg^2 &\Leftrightarrow \bar{n}_\mathrm{FRB,lin}=2.3\deg^{-2}\times\left(\frac{\sigma_{D,\mathrm{true}}}{\sigma_{D,\mathrm{fid}}}\right)^2 \\
    N^{DD}_{\ell,\rm EFT} = 1700\, \left(\frac{\mathrm{pc}}{\mathrm{cm}^3}\right)^2 \deg^2 &\Leftrightarrow \bar{n}_\mathrm{FRB,EFT}=5.8\deg^{-2}\times\left(\frac{\sigma_{D,\mathrm{true}}}{\sigma_{D,\mathrm{fid}}}\right)^2
\end{align}
at $z=0.2$, and
\begin{align}
    N^{DD}_{\ell,\rm lin} = 770\, \left(\frac{\mathrm{pc}}{\mathrm{cm}^3}\right)^2 \deg^2 &\Leftrightarrow \bar{n}_\mathrm{FRB,lin}=13\deg^{-2}\times\left(\frac{\sigma_{D,\mathrm{true}}}{\sigma_{D,\mathrm{fid}}}\right)^2 \\
    N^{DD}_{\ell,\rm EFT} = 300\, \left(\frac{\mathrm{pc}}{\mathrm{cm}^3}\right)^2 \deg^2
    &\Leftrightarrow \bar{n}_\mathrm{FRB,EFT}=33\deg^{-2}\times\left(\frac{\sigma_{D,\mathrm{true}}}{\sigma_{D,\mathrm{fid}}}\right)^2
\end{align}
at $z=0.5$, where $\sigma_{D,\rm fid}=100\,\mathrm{pc\,cm^{-3}}$ is the fiducial scatter used for Fig.~\ref{fig:snr} and $\sigma_{D,\rm true}$ is the true scatter. Thus, a $\sim10\%$ difference between $\sigma_{fid}$ and $\sigma_{D,\rm true}$ may modulate the density required for $C_\ell^{DD}$ SNR equilibrium by $\sim20\%$. 
Nonetheless, the signal domination up to $k\sim0.3\,h\,\mathrm{Mpc}^{-1}$ at $z=0.2$ achieved near the end of near-future surveys in the fiducial $\sigma_D$ estimate requires modeling past linear theory ($k\lesssim0.1\,h\,\mathrm{Mpc}^{-1}$), to one-loop theory (\S\ref{sec:formalism}) and potentially Hybrid Effective Field Theory (\S\ref{sec:heft}) for the auto-spectrum $P_{ee}$ ($C_\ell^{DD}$). The cross-spectrum $P_{eg}$ ($C_\ell^{Dg}$) will require these models earlier, due to the galaxy SNR exceeding one up to $k\sim0.5\,h\,\mathrm{Mpc}^{-1}$ for currently available samples. 

\section{Effective Field Theory}
\label{sec:formalism}

The description of FRB-galaxy cross-correlation observables ($C_\ell^{DD}$, $C_\ell^{Dg}$, $C_\ell^{gg}$) in terms of the respective power spectra ($P_{ee}$, $P_{eg}$, $P_{gg}$), and forecasts shown in Fig.~\ref{fig:snr} indicate the necessity to model free electrons to quasi-linear scales, on the same footing as galaxies. 
The concept of modeling the electrons as a linearly biased field of the matter density field ($\delta_e=b_e\delta_m$) has been suggested in the past (e.g.~\cite{Masui15}). 
Recently, refs.~\cite{Zhou25b,Andrew26} has confirmed this using hydrodynamical simulations (\S\ref{sec:sim}), measuring linear biases close to matter ($b_e\approx 1$). 
This model is standard in large-scale structure cosmology \cite{Kaiser84,Desjacques18}, where one must translate the observation of galaxy overdensity fields $\delta_g$ to constraints on $\delta_m$, as the latter encodes cosmological information. However, recent key analyses of LSS cosmology extend large-scale structure modeling further to smaller scales, using the effective field theory (EFT) one-loop power spectrum. Here, we will perform this extension for FRB-galaxy cross-correlations following the latest LSS power spectrum models \cite{Ivanov22b}. This will allow us to model the FRB-galaxy cross-clustering at higher precision and to significantly smaller scales.

The particular value of bias found by refs.~\cite{Zhou25b,Andrew26} confirms that the free electrons are nearly unbiased tracers of matter \cite{Masui15} at large (linear) scales. Using one-loop EFT and simulation measurements, we will extend this to significantly smaller scales, in \S\ref{sec:sim}. The (nearly) unbiased feature of free electrons benefits us in two aspects. One is the modeling extent of theory models, as highly-biased tracers are more difficult to model. Another is the ability to use FRBs to constrain baryonic feedback of galaxies, which is not only of interest to galaxy formation science, but also a major systematic in weak lensing surveys, such as LSST \cite{LSST} and Euclid \cite{Euclid}. This measurement is complementary to the kinetic Sunyaev-Zel'dovich (kSZ) effect, which probes ionized gas through secondary perturbations in the CMB and suffers a degeneracy with bulk velocity bias $b_v$ at large-scales \cite{kSZ,Smith18,Madhavacheril19,Schaan21,Amodeo21,Bigwood24,Hadzhiyska25,RiedGuachalla25}. 
We will revisit these topics again in \S\ref{sec:heft} and \S\ref{sec:sim}.

\subsection{Auto-spectrum}

While our primary interest is the DM-galaxy cross-spectrum $P_{eg}$, let us start with the modeling of the DM auto-spectrum $P_{ee}$ for clarity. 

The cosmological perturbation theory is the standard language to describe the LSS power spectra. This stems from the ability to model the large, quasi-linear scale behavior of the matter field, by establishing the equations of motion through gravity and making assumptions about the dark matter as pressureless fluid (Eulerian perturbation theory, EPT) or collisionless particles (Lagrangian perturbation theory, LPT) \cite{Bernardeau02,VlaWhiAvi15,Ivanov22b}. The two dark matter assumptions yield equivalent results \cite{McQWhi16,Chen20a,Maus24} and here we will adopt the latter convention. 
Through the gravitational equations of motion, LPT predicts the distribution of matter as 
\begin{equation}
    1 + \delta_m(\mathbf{x}) = \int d^3q \, \delta_D\big(\mathbf{x} - \mathbf{q} - \boldsymbol{\Psi}(\mathbf{q})\big), \qquad
    (2\pi)^3 \delta_D(\mathbf{k}) + \tilde{\delta}_m(\mathbf{k}) = \int d^3q \, e^{i\mathbf{k}\cdot(\mathbf{q}+\boldsymbol{\Psi})}
\end{equation}
where the Lagrangian displacement $\boldsymbol{\Psi}$ denotes the displacement from initial position $\mathbf{q}$ to position $\mathbf{x}$ at time $t$, i.e.~$\mathbf{x}(\mathbf{q}, t) = \mathbf{q} + \boldsymbol{\Psi}(\mathbf{q}, t)$. This can be solved perturbatively $\boldsymbol{\Psi} = \boldsymbol{\Psi}^{(1)} + \boldsymbol{\Psi}^{(2)} + \boldsymbol{\Psi}^{(3)} + \ldots$, where the orders denote the power of the initial conditions $\delta_0(\mathbf{q})$. The solution at each order can be found, for example, in refs.~\cite{Mat08a,ZheFri14,RamBuc12}. 

The free electron overdensity, which is distinct from the matter field predicted above, can be described by including all contractions of underlying initial fields allowed by symmetry arguments. To one-loop order at large scales, this is 
\begin{equation}
    F(\mathbf{q}) = 1 + b_1^L \delta_{0}(\mathbf{q})
    + \frac{b_2^L}{2}\left(\delta_{0}^2(\mathbf{q}) - \langle \delta_{0}^2 \rangle\right)
    + b_s^L \left(s_{0}^2(\mathbf{q}) - \langle s_{0}^2 \rangle\right)
    + \frac{b_{\nabla^2}^L}{4}\left(\nabla^2 \delta_{0}(\mathbf{q}) - \langle \nabla^2 \delta_{0} \rangle\right)
    + \mathcal{E}(\mathbf{q})
    \label{eq:F_q}
\end{equation}
where $s_0^2=s_0^{ij}s_0^{ij}$ is the self-contraction of the shear field $s_{0}^{ij} = \left(\partial_i \partial_j / \partial^2 - \delta_{ij}^K / 3\right) \delta_{0}$\footnote{It is standard practice to set the third-order, Lagrangian bias in cosmological analyses to zero, which corresponds to the coevolution approximation assuming number-conservation of galaxies and dynamics identical to matter.  See refs.~\cite{Desjacques18,Chen20c,DESI24-V} for justification.}. This way, we have compensated for our lack of knowledge regarding the electron distributions by allowing for additional (and maximal) degrees of freedom, introduced in the form of (Lagrangian) biases $b^L_X$. 
The bias functional $F(\mathbf{q})$ enters the prediction of the electron field $\delta_e$ as
\begin{equation}
    1 + \delta_e(\mathbf{x}) = \int d^3q \, F(\mathbf{q}) \, \delta_D(\mathbf{x} - \mathbf{q} - \boldsymbol{\Psi})
    \label{eq:deltae}
\end{equation}
which is directly used to predict the power spectrum
\begin{equation}
    \langle \delta_e(\mathbf{k}) \delta_e(\mathbf{k}') \rangle = (2\pi)^3 \delta_D^{(3)}(\mathbf{k} + \mathbf{k}') P_{ee}(k)
\end{equation}
where the Dirac Delta $\delta^D$ denotes the translational symmetry of the density fields. 
It is noteworthy that the `derivative bias' $b_{\nabla^2}$ gives rise to small-scale, `counterterm' contributions $P_{ee}(k)\supset \alpha_{e} k^2 P_{\rm lin}$, while the stochastic field $\mathcal{E}$ gives rise to the shot noise term $P_{ee}(k)\supset N_{ee}$; we will use $\alpha_{e}$ and $N_{ee}$ as the parametrization convention in this work\footnote{In galaxy redshift surveys the Poisson prediction for the shot noise $N=1/\bar{n}$, for 3D density $\bar{n}$, is commonly quoted as a rough estimate \cite{Maus25}, as galaxies are formed through (approximately) Poisson processes. This does not apply to free electrons.}. 
While observation of electrons through the dispersion measure are restricted to real space, extending the modeling to redshift space, which are required for modeling galaxies in cosmological galaxy redshift surveys, would give rise to more contributions depending on $\mu$, the cosine of the angle with the line of sight. Included here is the Fingers-of-God (FoG) dilations, which dilute the small-scale information due to stochastic motion. We discuss the FoG of free electrons in Appendix \ref{sec:fog}.

The Lagrangian biases introduced in Eq.~\ref{eq:F_q} can be converted to Eulerian biases, which model the electron density field $\delta_e$ as a function of the matter field at time $t$ (as opposed to initial fields in the Lagrangian framework)
\begin{equation}
    \delta_e = b_{e,1}^E \delta + \frac{b_{e,2}^E}{2}\delta^2 + b_{e,s}^E s^2 \quad .
\end{equation}
The conversion is (see e.g.~\cite{Desjacques18})
\begin{equation}
b_1^E = 1 + b_1^L, \quad 
b_2^E = b_2^L + \frac{8}{21} b_1^L, \quad b_s^E = b_s^L - \frac{2}{7} b_1^L
\label{eq:bias}
\end{equation}
The linear Eulerian bias $b_1^E$, in particular, is commonly quoted in literature as ``the bias'' of a field (e.g.~$\delta_{\rm tracer} = b_1^E\,\delta + \cdots$). Hence, we produce results for $b_1$ in Eulerian basis unless explicitly stated otherwise. 

\subsection{Cross-spectrum}

The galaxy overdensity field $\delta_g$ can be described similarly to the free electron overdensity field, but with an independent set of bias parameters $\{b^L_{g,1},b^L_{g,2},b^L_{g,s}\}$, counterterm $\alpha_{g}$, and shot noise $N_{gg}$. With the biases as inputs to the bias functional $F(\mathbf{q})$ (Eq.~\ref{eq:F_q}), one obtains a prediction for the galaxy density field $\delta_g$. By taking its auto-correlation ($\expval{\delta_g\delta_g}$) and cross-correlation with the electron field ($\expval{\delta_e\delta_g}$), one obtains the galaxy power spectrum $P_{gg}$ and electron-galaxy cross-spectrum $P_{eg}$. Almost all parameters of the cross-spectrum may be predicted from the auto-spectrum parameters, including the counterterm contribution sourced by small-scale dynamics \cite{Mergulhao23,Ebina24a}. However, the stochastic, shot-noise contribution is a new degree of freedom $N_{eg}$. The counterterm and stochastic contributions to the cross-spectrum become
\begin{equation}
    P_{eg} \supset \frac{1}{2}\left(\frac{b_{g,1}^E}{b_{e,1}^E}\alpha_{e} + \frac{b_{e,1}^E}{b_{g,1}^E}\alpha_{g}\right) k^2 P_{\rm lin} + N_{eg}
\end{equation}
in addition to the large-scale perturbative solutions predicted by $b_1$, $b_2$, and $b_s$. For galaxy surveys, cross-spectrum shot noise is typically expected to be $N_{eg}\sim f_{\rm over}\sqrt{N_{ee}N_{gg}}$, where $f_{\rm over}$ is an order unity parameter indicating the overlap between host structure of the respective galaxy populations \cite{Ebina24a}. In this case, we expect the structural overlap between free electrons and galaxies to be small, and by extension $f_{\rm over}\ll1$. This is confirmed in \S\ref{sec:sim} using simulation fits; notably, $N_{eg}$ is consistent with 0 within $2\sigma$ (Table \ref{tab:eft}).

Thus, the DM-galaxy cross-spectrum allows us to probe large-scale structure cosmology on the same footing as ongoing galaxy surveys up to $\sim 2\%$ additional uncertainty in amplitude (due to $f_e$) \cite{Andrew26}, providing an independent test of $\Lambda$CDM with extra sensitivity on the baryon energy density $\Omega_b$. The extra sensitivity on the baryon density can be particularly interesting, as current cosmological constraints on $\Omega_b$ are dominated by CMB measurements \cite{PCP18} and BBN \cite{Cooke18}, both of which are early-universe measurements. Late-time measurements of $\Omega_b$ from $C_\ell^{Dg}$ will provide a new test of $\Lambda$CDM that is not available with the current cosmological probes. 

In the derivation above we have restricted expressions to those present in real space ($\mu=0$), as is relevant for the projection $C_\ell^{Dg}$ and $C_\ell^{DD}$. We can trivially extend expressions for $P_{ee}$ and $P_{eg}$ to redshift space if necessary using the formalism already extensively developed for galaxy clustering (e.g.~\cite{Chen20a}).

\section{Hybrid Effective Field Theory}
\label{sec:heft}

The cosmological perturbation theory formalism described in \S\ref{sec:formalism} is bound in accuracy by the perturbative order to which the Lagrangian displacement $\boldsymbol{\Psi}$ and bias functional $F(\mathbf{q})$ are resolved. Extending the perturbative order indefinitely is infeasible, in part because the number of free parameters in $F(\mathbf{q})$ grows rapidly. 
The displacement $\boldsymbol{\Psi}$, however, can be extended using simulations in what is called the Hybrid Effective Field Theory (HEFT) \cite{Modi20,Kokron21,Hadzhiyska21,Zennaro22,PellejeroIbanez23,DeRose23b,Zennaro23,Nicola24,Zhou25a,Bartlett26,PellejeroIbanez26} technique. 
This approach informs $\boldsymbol{\Psi}$ using N-body simulations and is most effective when the higher-order Lagrangian biases are small, and its advantage over traditional EFT is greatest at low redshift, where nonlinear matter evolution is strongest. As currently formulated the technique performs best for real-space, rather than redshift-space, clustering. 
These characteristics --- $|b_n^L|\ll 1$, $z\approx 0$, and real-space clustering --- all indicate that FRBs are the ideal candidate for HEFT. 

In HEFT, one employs N-body simulations for the Lagrangian displacement $\boldsymbol{\Psi}$ (Eq.~\ref{eq:deltae}). Due to the accuracy in simulating the evolution of matter, this exceeds the precision of the perturbative solution for $\boldsymbol{\Psi}$, while retaining the general and robust perturbative framework for the bias functional $F(\mathbf{q})$ and density field.
In this framework, one can consider the tracer density field as a linear summation of `advected' (time-evolved) fields $F(\mathbf{k})$ 
\begin{equation}
    \delta_t(\mathbf{k}, a) = \left( 1 - \frac{b_{\nabla^2}^L k^2}{4} \right) \delta_{cb}(\mathbf{k}) + \sum_X b_X^L F_X(\mathbf{k}) + \mathcal{E}(\mathbf{k})
\end{equation}
where $X\in\{1,2,s\}$ and $\mathcal{E}$ the stochastic operator. This can directly be used to predict the auto- and cross-spectrum between tracers. To utilize this for a wide range of cosmologies (as is necessary in cosmological inference), it is common to use an emulator, such as \texttt{Aemulus$\nu$}, in order to predict the basis spectra (to within 0.25\% for $z<1$) \cite{DeRose23b,Shen25}. This extends the small-scale limit of analysis from $k_{\rm max}\sim k_{\rm nl}\sim 0.2$--$0.3\,h\,\mathrm{Mpc}^{-1}$ to $k_{\rm max}\sim0.6\,h\,\mathrm{Mpc}^{-1}$, increasing the number of modes available for analysis by a factor of $2^2$ to $3^2$, while also extending the possibility for parameter degeneracy breaking \cite{Kokron21}.
This technique has been adopted for DESI--CMB lensing and DES galaxy--galaxy lensing cross-correlations, which trace the real-space cross-correlation between matter and galaxies \cite{Hadzhiyska21,Sailer24,Chen24b,Maus25b}.

\begin{figure}
    \centering
    \includegraphics[width=\linewidth]{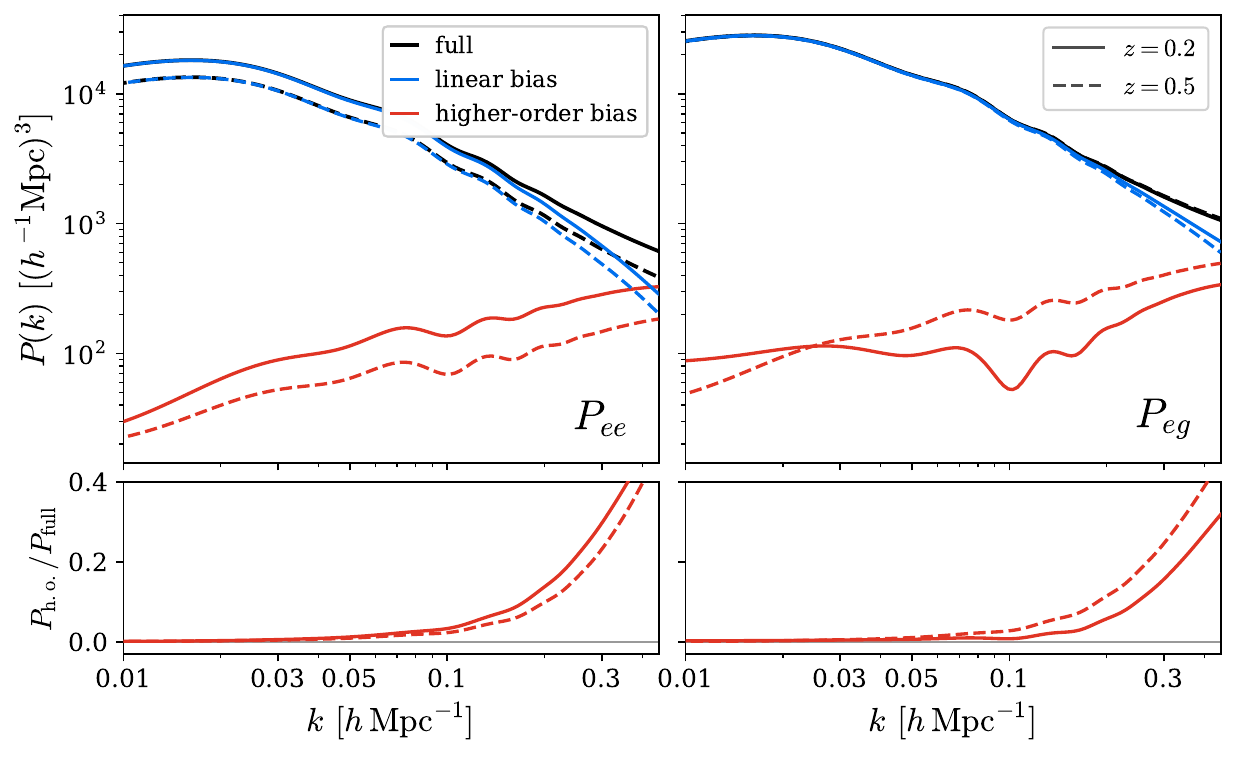}
    \caption{The free electron auto-spectrum $P_{ee}$ (left) and electron-galaxy cross-spectrum $P_{eg}$ (right) predicted by the best-fit EFT solution to the fiducial (\texttt{L1\_m9}) FLAMINGO simulation. The total spectrum is shown in black, the linear contribution (i.e.~that directly predicted by only $b_1$ and $N$) in blue, and the higher-order bias terms (with dependencies on $b_2$, $b_s$, and $\alpha$) shown in red. The top panels show the spectra comparison directly, while the bottom panels show the fractional contribution of higher-order bias terms. 
    In all cases, we find domination by linear bias terms to the extent of the EFT regime $k_\mathrm{nl}\sim0.2$--$0.3\,h\,\mathrm{Mpc}^{-1}$. 
    The strong electron-matter correlation as seen in Fig.~\ref{fig:rem} can be understood as a consequence of these weak nonlinear biases.}
    \label{fig:Pee_Peh_decomp}
\end{figure}

Due to the (relative) difficulty of simulating higher-order advected operators (e.g.~$F_2$ and $F_s$), the increased nonlinear matter evolution at low $z$, and ease of simulating real-space operators, this approach is most useful precisely for tracers like FRBs. 
Thus, it is instructive to consider the higher-order (one-loop) contribution to the linear-order contribution of $P_{ee}$ and $P_{eg}$, as shown in Fig.~\ref{fig:Pee_Peh_decomp} using the simulation fits from \S\ref{sec:sim}.
In both $z=0.2$ and 0.5, linear-bias terms dominate up to $k\sim0.5\,h\,\mathrm{Mpc}^{-1}$, far exceeding the typical linear regime $k\lesssim 0.1\,h\,\mathrm{Mpc}^{-1}$, and confirming the small influence of higher-order biases for electron spectra and its suitability as a HEFT application. 
Furthermore, in the absence of HEFT, these results indicate that for $\bar{n}_\mathrm{FRB}<10\deg^{-2}$ and $\sigma_D\approx100\,\mathrm{pc}\,\mathrm{cm}^{-3}$, the higher-order bias contributions are subdominant to observational noise for the EFT regime and can likely be marginalized using informative priors, e.g.~$b_X^L\sim 0$ as we will see in \S\ref{sec:sim}. 

\section{Validation with Hydrodynamical Simulations}
\label{sec:sim}

To test the one-loop modeling of $P_{ee}$ and $P_{eg}$ on a quasi-realistic, fully non-linear distribution of electrons we use the FLAMINGO suite of cosmological hydrodynamic simulations\footnote{\url{https://flamingo.strw.leidenuniv.nl/}} \cite{Schaye23,Kugel23,Helly26}. FLAMINGO evolves gas, dark matter, neutrinos, and stars with calibrated subgrid models for star formation and stellar/AGN feedback, and crucially provides a large library of feedback variants run from the same initial conditions, allowing us to assess how robust and sensitive the electron bias parameters are to the (uncertain) baryonic feedback prescription. We use the $1\,\mathrm{Gpc}$, fiducial-mass resolution box (\texttt{L1\_m9}, $1800^3$ gas particles, $h=0.681$, $\Omega_m=0.3046$) and its variants, at two redshifts, $z=0.2$ and $z=0.5$. There are eleven variants in total, nine of which are feedback variants, one (\texttt{NoCooling}) with no feedback or star formation, and one (\texttt{extraDM}) with finer dark matter particles.
Note that the \texttt{NoCooling} simulation is considered unrealistic, but we display it for comparison to visualize the magnitude of feedback.

For each gas particle we assign a free-electron weight $w_i = n_{e,i}\,m_i/\rho_i$, the number of free electrons it represents (here $n_e$ is the free-electron number density and $m_i/\rho_i$ the particle volume). By construction, star-forming particles carry $n_e=0$ in FLAMINGO, restricting the electrons to those outside structure. The weighted particles are deposited onto a $512^3$ mesh with a triangular-shaped-cloud (TSC) assignment and interlacing, and the free-electron overdensity power spectrum is estimated with \texttt{pypower}\footnote{\url{https://github.com/cosmodesi/pypower}} \cite{Hand17}, a part of the official DESI analysis pipeline. 

To mimic the galaxy populations most relevant for cross-correlation with FRBs, we build samples from the (sub-)halo catalogs constructed using the halo finder HBT-HERONS \cite{Forouhar25} and halo processor SOAP (Spherical Overdensity and Aperture Processor) \cite{McGibbon25} that approximate the DESI Bright Galaxy Survey (BGS)\footnote{Only a subset of BGS galaxies have been used for cosmological inference in DESI analyses to date. Here we intend to mimic the subset used for the latest DESI cosmology results, which amounts to $\bar{n}\simeq10^{-3}\,h^3\,\mathrm{Mpc}^{-3}$ \cite{DESI-DR2}, but the precise galaxy specification of either BGS or LRG should not affect the validity of the model.} at $z=0.2$ and the Luminous Red Galaxy (LRG) sample at $z=0.5$. We select central and satellite halos above a peak-mass threshold chosen to reproduce the target linear bias of each survey: $M>10^{12.5}\,M_\odot/h$ at $z=0.2$, giving $b_{g,1}\simeq1.45$, close to the DESI BGS value, and $M>10^{12.8}\,M_\odot/h$ at $z=0.5$, giving $b_{g,1}\simeq1.96$, DESI-LRG-like. Reassuringly, these bias-matched cuts also roughly reproduce the target number densities without much further tuning; the $z=0.5$ selection yields $\bar n\simeq4.6\times10^{-4}\,h^3\,\mathrm{Mpc}^{-3}$, comparable to the DESI LRG density, and at $z=0.2$ we obtain $\bar n\simeq10^{-3}\,h^3\,\mathrm{Mpc}^{-3}$ after random subsampling by $\sim80\%$. 
For each box and redshift we measure the three spectra entering the analysis: the electron auto-spectrum $P_{ee}$, the tracer auto-spectrum $P_{gg}$, and the electron--tracer cross-spectrum $P_{eg}$.

\begin{figure}
    \centering
    \includegraphics[width=\linewidth]{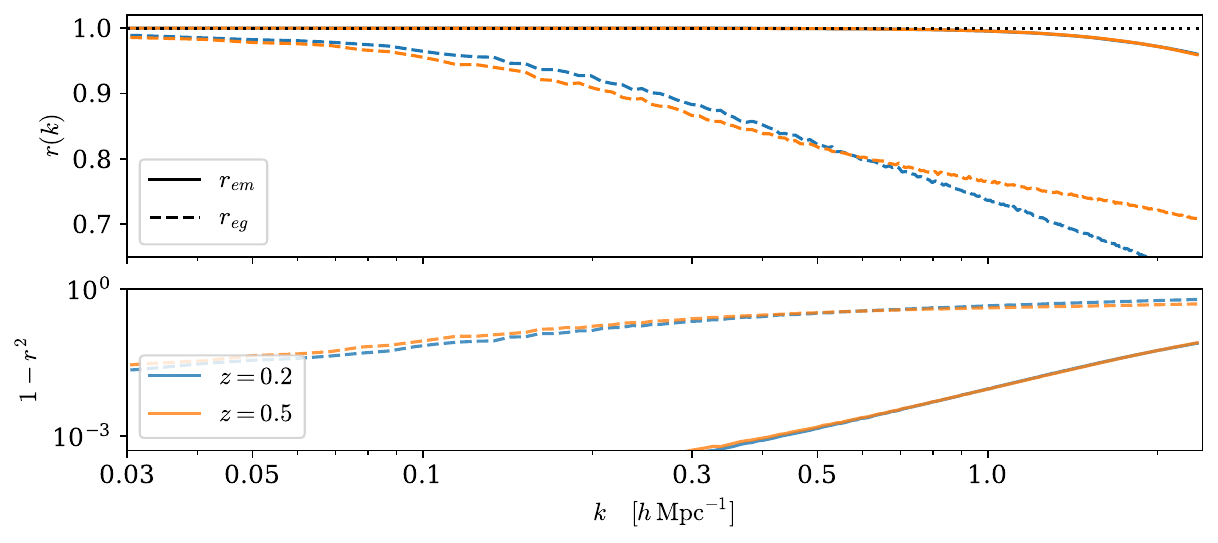}
    \caption{The correlation between free electrons and matter ($r_{em}$, solid), and between free electrons and galaxies ($r_{eg}$, dashed) from the fiducial (\texttt{L1\_m9}) FLAMINGO simulation. The top panel shows the correlation $r$ directly, whereas the bottom panel shows $1-r^2$, which informs the fractional power spectrum error from linear regression of the two fields. The correlation $r_{eX}$ is computed as the ratio between the cross-spectrum, $P_{eX}$, and geometric mean of the auto-spectra, $P_{ee}$ and $P_{XX}$ (for $X=\{m\,,g\}$). The electron-matter correlation is near-unity for sufficiently large-scale modes, throughout the entire quasi-linear regime ($k\lesssim 0.5 \,h\,\mathrm{Mpc}^{-1}$), with slight decorrelation at smaller scales. The decorrelation at small scales ($k\gtrsim 1 \,h\,\mathrm{Mpc}^{-1}$) is inevitable, as e.g.\ free electrons are underrepresented in overdense regions. 
    The electron-galaxy correlation is considerably weaker due to the significant bias and shot noise of the galaxies, although at linear-theory scales ($k\lesssim 0.1\,h\,\mathrm{Mpc}^{-1}$) the correlation is close to 1, as expected. The trend downwards to small scales is anticipated from the introduction of nonlinear effects. 
    }
    \label{fig:rem}
\end{figure}

In addition to the electrons and galaxies, we utilize the matter catalog to compute the electron-matter correlation through the auto-spectra $P_{ee}$ and $P_{mm}$, and the cross-spectrum $P_{em}$; i.e.~$r_{em}=P_{em}/\sqrt{P_{ee}P_{mm}}$. This, as seen in Fig.~\ref{fig:rem}, indicates nearly full correlation ($r_{em}\simeq1$) up to $k\sim1\,h\,\mathrm{Mpc}^{-1}$, with slight decorrelation beyond that.
This extremely high correlation is indicative of the free electrons in FLAMINGO being `unbiased' tracers of matter, as hinted in ref.~\cite{Masui15}, and verifies the (previously implicitly assumed) ability to constrain baryonic feedback through FRBs. 
The statement of $r_{em}\simeq1$ and small non-linear electron biases ($b_{n>1}^{L}\approx 0$) are equivalent up to the small-scale limit of the model \cite{Ebina26e}, and we will confirm this later again through perturbation theory fits, e.g.~in Table \ref{tab:eft} or Fig.~\ref{fig:Pee_Peh_decomp}. 
The electron-galaxy correlation $r_{eg}$ can be computed in an identical manner and is also shown in Fig.~\ref{fig:rem}. This shows the expected trajectory, where high correlation ($r\gtrsim 0.95$) is maintained at large scales and nonlinear effects degrade the correlation at quasi-linear scales or smaller ($k\gtrsim0.1\,h\,\mathrm{Mpc}^{-1}$). The difference in the behavior in the lower panel at low $k$ is largely attributable to shot noise in the galaxy field. 

We jointly fit $\{P_{ee}, P_{eg}, P_{gg}\}$ with the real-space one-loop model of Section~\ref{sec:formalism}, using the theory software \texttt{velocileptors}\footnote{\url{https://github.com/sfschen/velocileptors}} \cite{Chen20a,Chen20c} and holding the cosmology fixed to the simulation truth. The first set of free parameters are the two sets of nonlinear bias parameters $\{b_{e,1},b_{e,2},b_{e,s}\}$ and $\{b_{g,1},b_{g,2},b_{g,s}\}$, which we sample directly. 
Most remaining parameters, the counterterms $\alpha_{e},\alpha_{g}$ and the shot noise $N_{gg},N_{eg}$, enter the model linearly, and we therefore marginalize them by analytic marginalization, integrating over them with conservative Gaussian priors so that only the nonlinear biases are sampled; we refer the reader to refs.~\cite{Sailer24,Maus25} for description and validation of analytical marginalization. Finally, we directly marginalize over $N_{ee}$ in order to enforce the physical lower bound that $N_{ee}>0$ by sampling with a wide, flat prior.
We adopt a Gaussian covariance appropriate to the simulation box volume. The fit range is $k<k_{\rm max}$, with $k_{\rm max}=0.20\,h\,\mathrm{Mpc}^{-1}$ at $z=0.2$ and $0.22\,h\,\mathrm{Mpc}^{-1}$ at $z=0.5$. This matches the small-scale limit predicted from EFT ($k_{\rm nl}$ \cite{Bernardeau02}; see e.g.~Fig.~1 of ref.~\cite{Sailer24}) and slightly exceeds the small-scale limit employed in DESI key analysis \cite{DESI24-V}. The inference is conducted using a Markov Chain Monte Carlo (MCMC) with the \texttt{cobaya} software \cite{Cobaya,CobayaCode}. 

\begin{figure}[t]
    \centering
    \includegraphics[width=\textwidth]{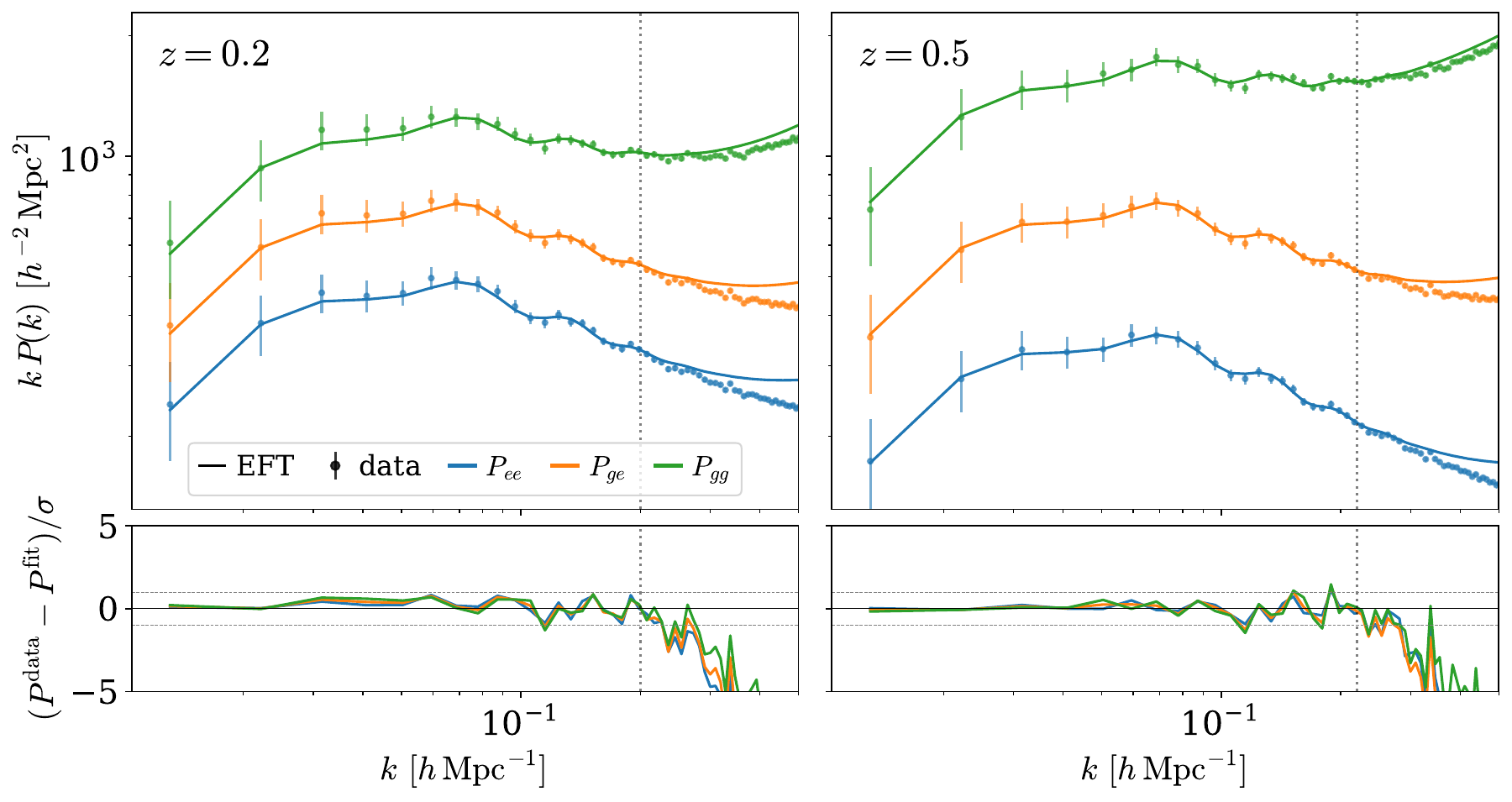}
    \caption{Best-fit one-loop EFT model (curves) compared to the FLAMINGO fiducial-box measurements (points) of the electron auto-spectrum $P_{ee}$, the electron-tracer cross-spectrum $P_{ge}$, and the tracer auto-spectrum $P_{gg}$, plotted as $k\,P(k)$. The lower panels show the residuals in units of the measurement error. The model is fit jointly to all three spectra below $k_{\rm max}$ (denoted by vertical lines) and describes them within noise across the fit range, with reduced-$\chi^2=43.2/(63-11)$ at $z=0.2$ and $38.0/(69-11)$ at $z=0.5$. Curves past $k_{\rm max}$ (dotted, black lines) are an extrapolation. }
    \label{fig:bestfit}
\end{figure}

The one-loop model provides an excellent description of all three spectra over the full fit range.  For the fiducial box we find reduced $\chi^2 = 43.2/(63-11)$ at $z=0.2$ and reduced $\chi^2 = 38.0/(69-11)$ at $z=0.5$, with best-fit residuals consistent with noise out to $k_{\rm max}$ for $P_{ee}$, $P_{eg}$, and $P_{gg}$ simultaneously (Figure~\ref{fig:bestfit}). The recovered large-scale electron bias is $b_{e,1}\simeq0.92$ at both redshifts, in agreement with linear theory fits \cite{Zhou25b,Andrew26}, while the tracer biases recover the targeted $b_{g,1}\simeq1.45$
($z=0.2$) and $1.95$ ($z=0.5$). The higher-order biases further confirm that the electrons are unbiased tracers of matter, with both $b_2^L$ and $b_s^L$ consistent with zero to $<1.5\sigma$ (hence consistent with the coevolution approximation as well). Our fits are also consistent (to $< 1.5\sigma$) with the peak-background split estimates of the biases \cite{Cole89,Sheth99,Whi14}, where $b_n^L\sim (b_1^L)^n$.

\begin{table}
    \centering
    \caption{EFT parameter fits for the free electrons in the fiducial (\texttt{L1\_m9}) simulation. Simulation model variants do not affect the parameter fits, as shown in Figs.~\ref{fig:contour1}, \ref{fig:contour2}. All biases are Lagrangian, each of which can be converted to Eulerian by the conversion shown in Eq.~\ref{eq:bias}. $\alpha_{e}$ is in $h^{-2}{\rm Mpc}^2$,
    $N_{ee}$, $N_{eg}$ are in $h^{-3}{\rm Mpc}^3$. }
    \label{tab:eft}
    \begin{tabular}{lcccccc}
    \hline
    $z$ & $b_1^L = b_1^E - 1$ & $b_2^L$ & $b_s^L$ & $\alpha_{e}$ & $N_{ee}$ & $N_{eg}$ \\
    \hline
    $0.2$ & $-0.075 \pm 0.023$ & $-0.55 \pm 0.43$ & $0.62 \pm 0.53$ & $2.6 \pm 4.2$ & $255 \pm 216$ & $526 \pm 342$ \\
    $0.5$ & $-0.082 \pm 0.022$ & $-0.52 \pm 0.41$ & $0.58 \pm 0.56$ & $-1.1 \pm 4.3$ & $237 \pm 191$ & $568 \pm 419$ \\
    \hline
    \end{tabular}
\end{table}

\begin{figure}[t]
    \centering
    \includegraphics[width=\textwidth]{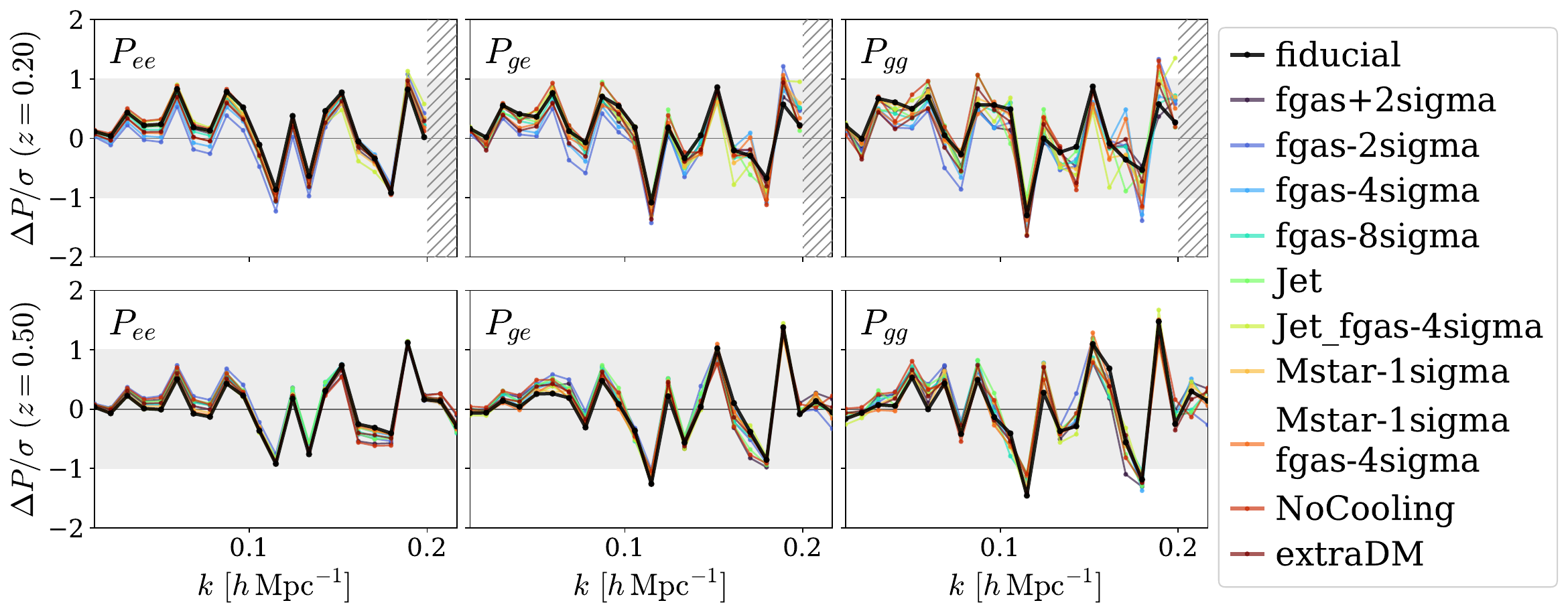}
    \caption{Best-fit residuals for each of the three spectra ($P_{ee}$, $P_{ge}$, $P_{gg}$, from left to right) at $z=0.2$ (top) and $z=0.5$ (bottom) bins, across all eleven FLAMINGO model variants. The hashed region on the top row indicates the region beyond $k_\mathrm{max}=0.2\,h\,\mathrm{Mpc}^{-1}$ at $z=0.2$. The deviations stay within $2\sigma$ over the fit range for every variant, demonstrating that the one-loop model fits the electron and cross spectra equally well regardless of the baryonic feedback prescription. The errors ($\sigma$) are calculated using the Gaussian (disconnected) approximation and thus may be underestimated.
    }
    \label{fig:residual}
\end{figure}

\begin{figure}
    \centering
    \includegraphics[width=\linewidth]{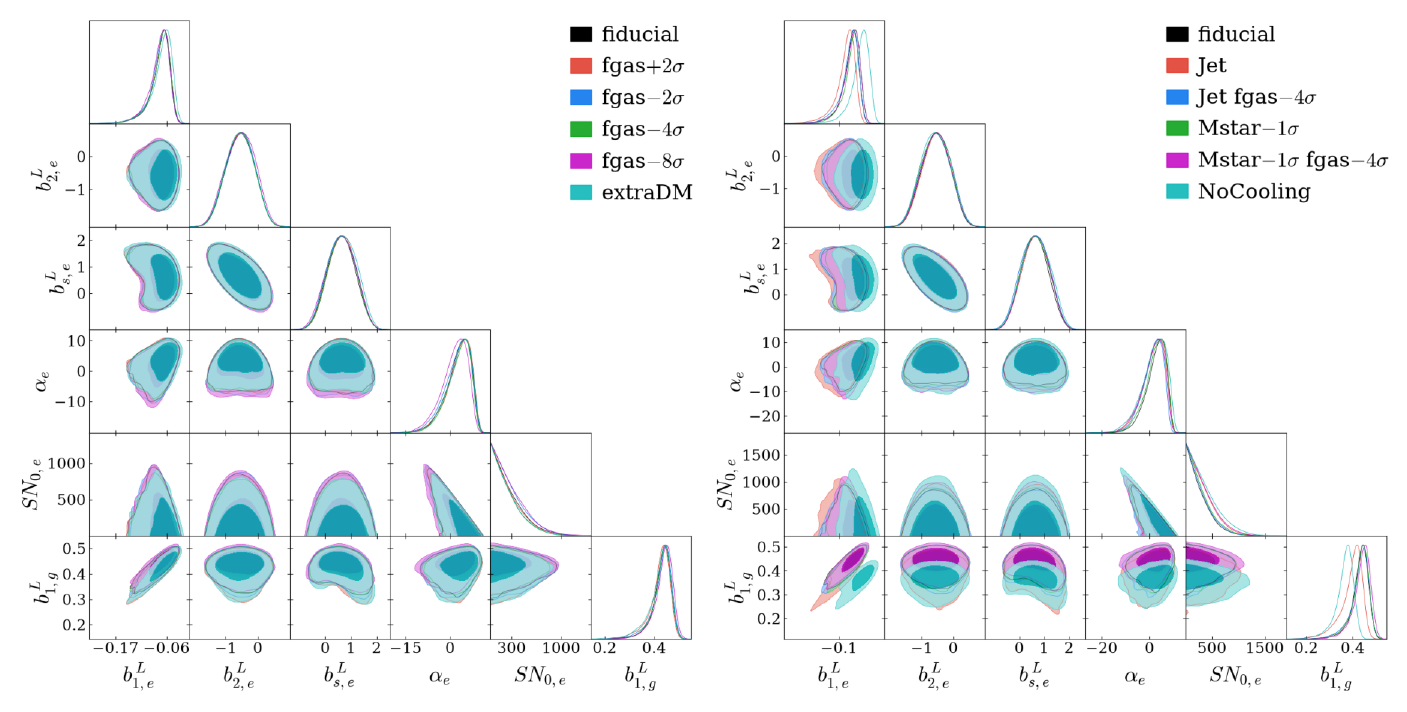}
    \caption{The marginalized posterior for a subset of EFT nuisance parameters used to jointly model the $P_{ee}$, $P_{eg}$, and $P_{gg}$ spectra at $z=0.2$, for DESI BGS-like samples. The posteriors indicate that the EFT parameters are insensitive to feedback model variations, with the exception of the \texttt{NoCooling} model, which is the (unrealistic) scenario with no feedback or star formation. The biases are all Lagrangian.}
    \label{fig:contour1}
\end{figure}

\begin{figure}
    \centering
    \includegraphics[width=\linewidth]{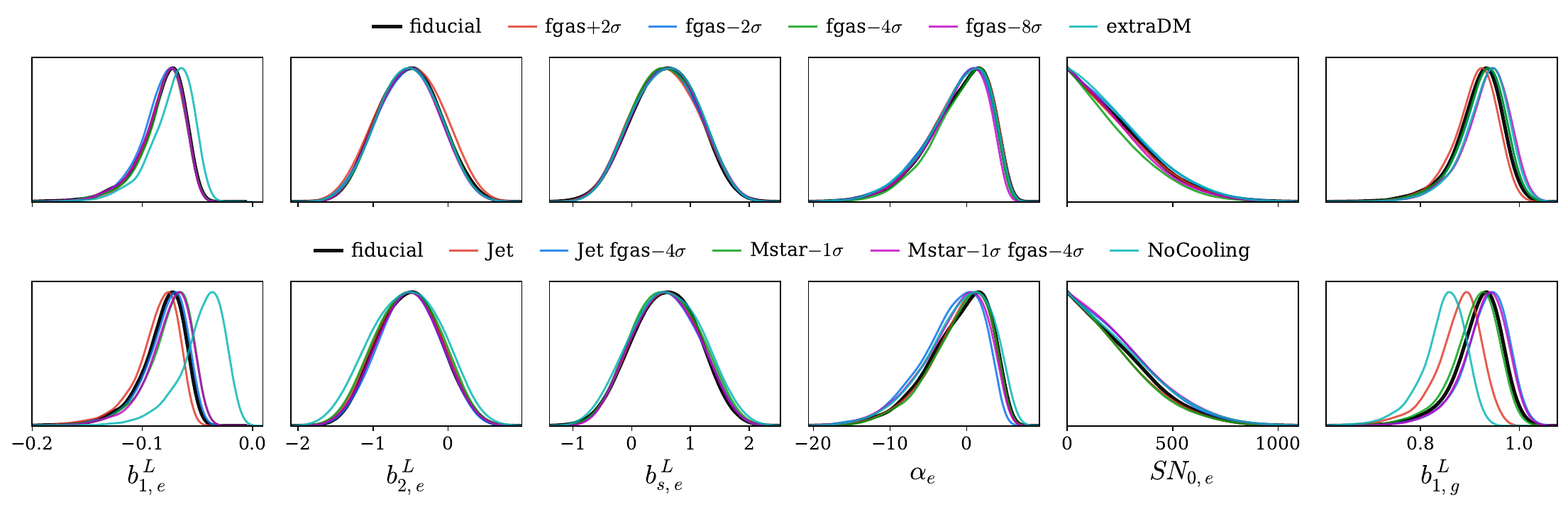}
    \caption{1D posteriors obtained as a result of MCMC inference over $P_{ee}$, $P_{eg}$, and $P_{gg}$ spectra at $z=0.5$, with DESI LRG-like samples. As the 2D contour degeneracies are qualitatively similar to Fig.~\ref{fig:contour1} at $z=0.2$, we choose to focus on the 1D posterior here. }
    \label{fig:contour2}
\end{figure}

To gauge the impact of baryonic feedback we repeat the measurement and fit for eleven FLAMINGO simulation variants spanning the calibrated feedback range: the four $f_{\rm gas}{\pm}$ excursions, jet-mode AGN, stellar-mass-shifted models, a no-cooling extreme, and others. Across this entire suite the best-fit residuals remain below $\sim2\sigma$ within the fit range (Fig.~\ref{fig:residual}), with all electron nuisance parameters stable against feedback models (Figs.~\ref{fig:contour1} and \ref{fig:contour2}), with the exception of the \texttt{NoCooling} model shifting $b_{e,1}$ by $\sim1\sigma$ (to $\simeq0.98$, as \texttt{NoCooling} reduces baryon-specific behavior). This confirms that the EFT modeling, including the cross-spectrum $P_{eg}$, is robust to feedback model variations, with only small shifts in the EFT parameters from model to model. This agrees with and further extends the consistency of linear bias $b_{1,e}$ found in refs.~\cite{Zhou25b,Andrew26} to all one-loop EFT parameters, and reinforces the idea that dispersion measures are a nearly unbiased tracer of matter \cite{Masui15} with further model sophistication, including higher-order biases. 

The highly correlated residuals across the different FLAMINGO model variants seen in Fig.~\ref{fig:residual} suggest that we are seeing noise from the particular realization of large-scale structure in the box, rather than limitations of the model. This would imply that larger volume simulations should have improved agreement. Alternatively, one could employ control variates techniques (e.g.\ refs.~\cite{Kokron22,Hadzhiyska23,DeRose23a,Bartlett26}) to achieve some of the same noise reduction without further resources.

\section{Conclusion}
\label{sec:conclusion}

Fast radio burst (FRB) dispersion measures offer a window into the late-time distribution of baryons. This, combined with the rapidly expanding experiment capabilities (as shown in e.g.~Fig.~\ref{fig:snr}), allows us to pursue multiple science cases in the near future. Two among these are probing baryonic feedback \cite{Ebina26e}, one of the leading systematics for Stage-IV weak lensing surveys, such as LSST \cite{LSST} and Euclid \cite{Euclid}; and large-scale structure cosmology, by modeling the electrons probed through FRBs on the same footing as other tracers of matter, like galaxies. In this work, we employ both effective field theory (EFT) and the FLAMINGO hydrodynamical simulations \cite{Kugel23,Schaye23,Helly26} to further prepare for these science goals. 

It has been suggested (e.g.~\cite{Masui15}) and verified using hydrodynamical simulations \cite{Zhou25b,Andrew26} that the linear bias prescription $\delta_e=b_e \delta_m$ is sufficient to model the distribution of electrons at large scales. Here, we further extend this using the full EFT prescription used in key analysis of galaxy surveys \cite{DESI24-V}. This not only places the FRB dispersion theory on the same footing as that of current large-scale structure cosmology, but also significantly expands the modeling extent and thereby allows for drastically more Fourier modes to be included in the analysis. With the acceleration of FRB data collection, it is expected that these theories will be necessary to utilize all signal-dominated modes of FRBs during the operation of next-generation experiments. 
We perform this model construction for both FRB dispersion auto-spectra and FRB-galaxy cross-spectra in \S\ref{sec:formalism}, and validate against outputs of FLAMINGO simulations in \S\ref{sec:sim}. We also assess the galaxy samples required for these cross-correlations, finding that current DESI-like spectroscopic samples provide the necessary redshift coverage and number density. With the FRB noise dominating the error budget, future gains will come primarily from the growth of FRB catalogs rather than galaxy catalogs. 

FRB-galaxy cross-correlations probe the free electron-galaxy cross-spectra. Thus,
to use $P_{eg}$ as a proxy for the matter–galaxy cross-spectrum and constrain baryonic feedback, one must establish that electrons trace matter closely. Ref.~\cite{Masui15} has suggested that electrons are unbiased tracers of matter, which refs.~\cite{Zhou25b,Andrew26} recently supported with linear bias measurements close to unity.
We further verify this from two directions, in \S\ref{sec:sim}. First, we observe the electron-matter cross-correlation directly from the simulation outputs in Fig.~\ref{fig:rem}, finding extremely high correlations ($r_{em}\approx1$) up to $k\sim1\,h\,\mathrm{Mpc}^{-1}$. Second, the electron-galaxy joint fit results using EFT yield small Lagrangian biases, which indicate that the electrons probe matter closely, up to quasi-linear scales described in the theory. 
These both pave the way for the use of FRBs to measure baryonic feedback. 
The latter also indicates that while data errors dominate, one can likely marginalize over higher-order biases using informative priors. 

The EFT formalism developed in this work can be expanded using Hybrid Effective Field Theory (HEFT). In particular, FRBs, which probe nearly unbiased low-redshift tracers in real space, are an ideal candidate for HEFT application. This would further expand the modeling extent from EFT models and we explore this direction in \S\ref{sec:heft}, though validation is left for future work. 
Beyond HEFT, it will be useful to repeat this test with a wider range of hydrodynamical simulations, whose predictions can differ given the complexity of the baryonic physics involved \cite{Hadzhiyska25,Bigwood26,Siegel26}.
Finally, the models developed here can be applied once next-generation survey data arrive, and will become necessary as the FRB catalogs grow.

\section*{Acknowledgements}

HE and MW were supported by the DOE. 
This research used resources of the National Energy Research Scientific Computing Center (NERSC), a Department of Energy User Facility.
Claude Code (Opus $\le5.5$ and Fable 5) was used for software implementation in this work. 
All writing and analysis choices were made by the authors, and the authors take full responsibility for the contents of this work.

\appendix
\section{Fingers of God}
\label{sec:fog}

\begin{figure}
    \centering
    \includegraphics[width=\linewidth]{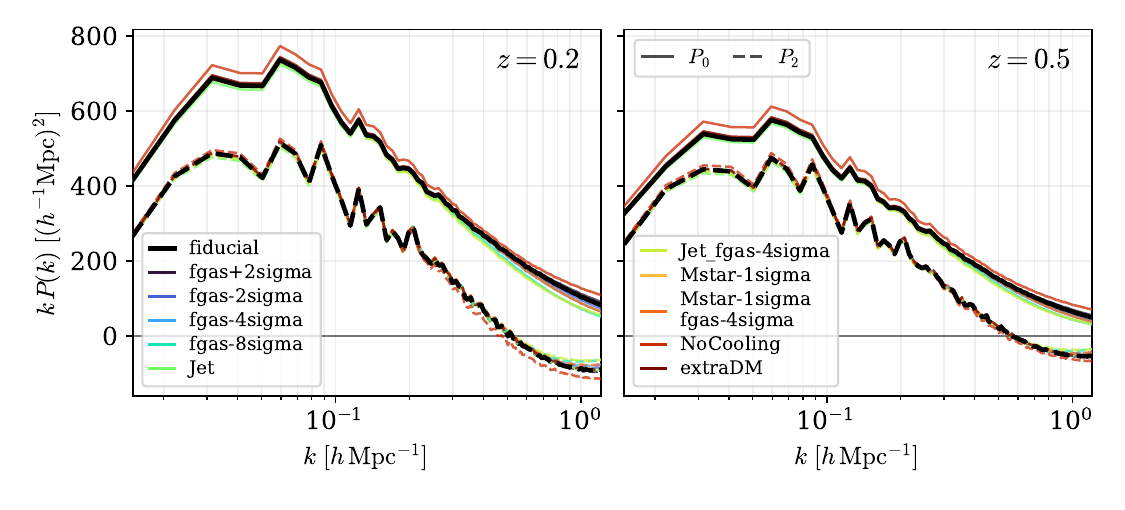}
    \caption{The free electron auto-spectrum monopole ($P_0$) and quadrupole ($P_2$) from the $z=0.2$ and 0.5 boxes of the FLAMINGO simulation. All models consistently indicate that the $P_2$ crosses zero at $k\gtrsim0.5\,h\,\mathrm{Mpc}^{-1}$, indicating weak FoG effects \cite{BaleatoLizancos25}.}
    \label{fig:multipoles}
\end{figure}

Although we do not anticipate measuring the clustering of free electrons in redshift space, the magnitude of their Fingers of God (FoG) effect is still an intriguing one. The FoG effect can be modeled as the stochastic motion of tracers (typically galaxies) with respect to matter \cite{Dodelson20} and damps the power spectrum along the line of sight. As free electrons exist outside of structure, one would expect that the random virial velocities are considerably smaller than galaxies that reside within halos. A simple heuristic that can assess this is the zero crossing of the power spectrum quadrupole \cite{BaleatoLizancos25}. Using the velocity data of the FLAMINGO simulations (\S\ref{sec:sim}), we measure the power spectrum monopole and quadrupole in redshift space, as shown in Fig.~\ref{fig:multipoles}. This demonstrates that the quadrupole zero crossing is $k\gtrsim0.5\,h\,\mathrm{Mpc}^{-1}$ for both redshifts. This is considerably higher than the galaxy samples considered in ref.~\cite{BaleatoLizancos25} and far exceeds the nonlinear limit $k_{\rm nl}\sim 0.2-0.3\,h\,\mathrm{Mpc}^{-1}$ of perturbative modeling at these redshifts; both indicate that the free electron FoG are weak, as we expect. 

\bibliographystyle{JHEP}
\bibliography{main}

\end{document}